\documentclass[]{interact}

\usepackage[normalem]{ulem}
\usepackage{xcolor}

\usepackage{comment}

\usepackage{lmodern}

\usepackage[authoryear]{natbib}

\usepackage{epstopdf}
\usepackage[caption=false]{subfig}

\theoremstyle{definition}

\theoremstyle{remark}

\newcommand\redsout{\bgroup\markoverwith{\textcolor{red}{\rule[0.5ex]{2pt}{0.4pt}}}\ULon}

\begin{document}


\title{Local Quasi-Exponential Growth Models: Kernel Differential Equation Regression and Mouse Tumor Growth Data}


\author{
\name{W. John Braun$^\ast$\textsuperscript{a}\thanks{$^\ast$CONTACT W. John Braun. Email: john.braun@ubc.ca} and Chunlei Ge\textsuperscript{a}}
\affil{\textsuperscript{a}The University of British Columbia Okanagan, 3333 University Way, Kelowna, BC, Canada}
}

\maketitle

\begin{abstract}
Local polynomial regression faces several challenges when dealing with sparse data. The difficulty in capturing local features of the underlying function can lead to a possible misrepresentation of the true relationship. Furthermore, with limited data points in local neighborhoods, the variance of estimators can increase significantly. Local polynomial regression also requires a substantial amount of data to produce good models, making it less efficient for sparse datasets. This paper employs a differential equation-constrained regression approach, introduced by \citet{ding2014estimation}, for local quasi-exponential growth models. By incorporating first-order differential equations, this method extends the sparse design capacity of local polynomial regression while reducing bias and variance. We discuss the asymptotic biases and variances of kernel estimators using first-degree Taylor polynomials. Model comparisons are conducted using mouse tumor growth data, along with simulation studies that include tumor growth with different sparse designs, and simulated quasi-exponential growth with varying levels of noise and growth rates.
\end{abstract}


\section{Introduction}
Suppose $n$ independent paired random variables $\{X_i, Y_i\}_{i=1}^n$ are observed, where it is assumed that
\begin{equation}\label{basemodel} Y_i = m(X_i) + \sigma(X_i) \epsilon_i, \ \ \ \ \ i = 1, \ldots, n. \end{equation}
The noise terms $\epsilon_i$ are assumed to be mean-zero random variables with 
unit variance, independent of the covariates $X_i$, which may be distributed according
to a random or a fixed design on an interval $(a, b)$.  In both cases, our interest will center on local polynomial estimation of the regression function
\[ m(x) = E[Y|X = x]. \]
The function $m(x)$ is usually assumed to have a certain degree of smoothness as
is the function $\sigma(x)$.  Specific conditions in the context of local linear regression are laid out, for example, in the paper by \citet{cheng2018bias}.   
\citet{fan1996local} and  \citet{wand1994kernel} are classic references in the area of local polynomial regression which is designed to nonparametrically 
 estimate $m(x)$, for any evaluation point $x \in (a, b)$. 

For each evaluation point $x$, the local $p$th degree polynomial regression estimator is obtained by minimizing
\begin{equation}\label{eqn:kernelReg} \sum_{i=1}^n \left(y_i -
\sum_{j=0}^p \beta_j(x_i-x)^j/j!\right)^2
K_h(x_i - x) \end{equation}
with respect to a set of polynomial coefficients $\beta_0, \beta_1, \ldots, \beta_p$.
The kernel is denoted by $K_h(u) = K(u/h)/h$,  where $K(x)$ is usually a
 symmetric probability density.  The scalar $h$ is referred to as the bandwidth.

By identifying the expression $\sum_{j=0}^p \beta_j(x_i-x)^j/j!$ with the Taylor
polynomial of $m(x_i)$ about $x$, we have $\beta_j = m^{(j)}(x)$, the 
$j$th order derivative of the function $m(x)$.   Thus, the estimator of $m(x)$ is 
the $\widehat{\beta}_0$ obtained from minimizing (\ref{eqn:kernelReg}).  

Local polynomial regression avoids restrictive assumptions about functional forms or error distributions, making it suitable in scenarios where parametric models are
unknown or may fail.  Local regression is often capable of capturing complex, nonlinear relationships that parametric methods might miss. \citet{fan1996local} explored the asymptotic properties of local polynomial regression, while \citet{ruppert1994multivariate} developed an asymptotic distribution theory for multivariate local regression.

 Traditionally, local polynomial regression struggles with sparse data regions, especially at boundaries. By incorporating differential equation constraints, local polynomial regression can better interpolate in sparse regions, reducing bias and variance. In this paper, we will employ a differential equation-constrained regression approach, introduced by \citet{ding2014estimation}, for the local quasi-exponential growth model, which is a relatively simple but general growth model.

 Incorporating information from differential equations can help to alleviate aforementioned limitations in an effective way. The work of \citet{ding2014estimation} primarily addresses parameter estimation for differential equations, while also introducing a differential equation–constrained local polynomial regression framework. 
 \citet{su2013local} used local polynomial regression to solve boundary value problems for second order differential equations; while their approach bears a superficial 
 resemblance to the methodology to be studied in the present paper, they do not
 consider the statistical modeling of observational data. 
 A closely related one-step parameter estimation approach for differential equations was independently proposed by \citet{hall2014quick}. Our previous work, \citet{ge2026differential}, discussed a local exponential growth model, which is constrained by an exponential-type differential equation. This paper will extend that work through the exploration of a quasi-exponential type differential equation.

 This paper is organized as follows. Section 2 outlines the DE-constrained local polynomial regression method. Section 3 discusses the asymptotic properties of DE-constrained estimation. The numerical properties are discussed in Section 4 with two application examples to mouse growth data. After that, there is a discussion section to review our method and outline our future work.



\section{Deterministic Exponential and Generalized Growth Models}


Perhaps the simplest differential equation-based model is that for exponential growth: for $t > t_0$, where $g(t)$ is a real-valued function satisfying
\[ g'(t) = \lambda g(t), \ \ \ \ g(t_0) = g_0.  \]
Here, $t_0$ is the initial observation time, $g_0 > 0$ is the size at time $t_0$, and 
$\lambda$ is the growth rate parameter.  

The exponential model often leads to overestimates of growth at later times
\citep{chowell2016growing}.  Using a model described by \citet{tolle200387}, \citet{viboud2016generalized} developed the Generalized-Growth Model which they
successfully applied to a number of epidemic data sets: 
\begin{equation}\label{eq:GG} 
g'(t) = \lambda g^{\alpha}(t), \ \ \ \ g(t_0) = g_0.  \end{equation}
Here, $\alpha$ is referred to as deceleration parameter \citep{viboud2016generalized}, 
since when it is less than 1, growth will be subexponential, beginning with growth
which is nearly exponential but decelerating over time to polynomial-time growth.  

We note that it is also possible for that $\alpha \geq 1$.  Equality corresponds
to the exponential case, while strict inequality corresponds to a super-exponential
form of growth which can occur temporarily.  When $\alpha > 1$, it can be shown
that the initial value problem satisfies a Lipschitz condition when 
\[ t < b = t_0 + \frac{g_0^{1-\alpha}}{(\alpha-1)\lambda}. \]
Thus, a unique solution to the initial value problem can be found on the interval $[t_0, b]$.  Nevertheless, the important applications are usually in
the sub-exponential case.  

In the context of epidemics, \citet{viboud2016generalized}
argue that sub-exponential growth could be a result of clustering, 
changes in population behaviour and differences in susceptibility and
infectivity in the underlying population.  Analogous considerations 
could lead to such behaviour in growing tumours, particularly with respect
to the latter two points: tumour cell behaviour could change over time in response
to changes in their environment, and some cells may be more resistant to carcinogenic
effects than others.  

Since our ultimate goal is to model noisy real growth data, and such data are positive, a log transformation of the volume measurements may be
appropriate.   Therefore, we suggest that the Generalized Growth model might be better formulated as: for $t > t_0$, 
\begin{equation}    
G'(t) = \lambda e^{(\alpha-1)G(t)}, \ \ \ \ \ G(t_0) = G_0
\label{eqn:quasilog1}
\end{equation}
where $0<\alpha \leq 1$, $\lambda \neq 0$, and $G_0$ is the natural log of the size at $t_0$.  This model is
mathematically equivalent to the model (\ref{eq:GG}), with 
$G(t)=\log (g(t))$.  

\section{Statistical Models}

Having established the generalized exponential growth framework, we now develop statistical estimation procedures that leverage this structure within a local polynomial regression context.

Given observations $(t_1, y_1), \ldots, (t_n, y_n)$, we seek an estimate of
the regression function $g(t)$ which satisfies: 
\begin{equation}    
y_i = g(t_i) + \varepsilon_i,  \ \ \ \ i = 1, 2, \ldots, n.  
\label{eqn:quasi1}
\end{equation}
It is assumed that $g'(t) = \lambda g^{\alpha}(t)$, $0<\alpha\leq 1$, $\lambda \neq 0$, and the $\varepsilon_i$'s are  uncorrelated, mean-zero errors. The common variance of $\varepsilon$ is $\sigma_{\varepsilon}^2$. 

An alternative form of the model is based on a multiplicative error assumption: 
\begin{equation}    
y_i = g(t_i) e^{\epsilon_i},  \ \ \ \ i = 1, 2, \ldots, n. \nonumber
\end{equation}
After a natural log transformation, this model becomes
\begin{equation}\label{eq:Gmodel}    
y_i = G(t_i) + {\epsilon_i},  \ \ \ \ i = 1, 2, \ldots, n. 
\end{equation}
where $G(t)$ satisfies (\ref{eqn:quasilog1}).  

\subsection{Nonlinear Regression}
The explicit solution to the differential equation in model (\ref{eq:GG}) is given by
\begin{equation}    
g(t)=\{(1-\alpha)\lambda (t-t_0) +g(t_0)\}^{1/(1-\alpha)}.\label{eqn:soln}
\end{equation}
Therefore, the statistical problem posed at (\ref{eqn:quasi1}) could
be solved by applying nonlinear least-squares to estimate the parameters $\alpha$, $\lambda$ and $g(t_0)$:
\[ \min_{\alpha, \lambda, g(t_0)} \sum_{i=1}^n (y_i - g(t_i))^2.  \]
The form of the explicit solution suggests that difficulties could
arise when $\alpha$ is near the value 1, that is, in the region
of the parameter space where there would be a transition from
polynomial behaviour in the solution to exponential behaviour.  

In the log-scale case, the explicit solution to the differential equation in model (\ref{eqn:quasi1}) is given by
\begin{equation}    
G(t)=\frac{1}{1-\alpha}\log\left\{(1-\alpha)\lambda (t-t_0) +e^{(1-\alpha)G(t_0)}\right\}.  \label{eqn:solnG}
\end{equation}

Again,  the statistical problem posed at (\ref{eq:Gmodel}) could
be solved by applying nonlinear least-squares:
\[ \min_{\alpha, \lambda, G(t_0)} \sum_{i=1}^n (\log(y_i) - G(t_i))^2.  \]
The parameter estimates will be different from the ones obtained
using the original scale, but they will agree, in the large sample limit.  The issues with estimation of $\alpha$ in the transition region remain.  

\subsection{Differential Equation-Constrained Local Regression}

The alternative to parametric nonlinear regression that we consider is
differential equation-constrained local polynomial regression.  In view of
(\ref{eqn:kernelReg}) and the subsequent comments,  for a given evaluation 
point $x$, we
set down the general form of the least-squares objective as
\begin{equation}\label{eqn:localDEReg} \sum_{i=1}^n \left(y_i -
\sum_{j=0}^p m^{(j)}(x)(x_i-x)^j/j!\right)^2
K_h(t_i - t). \end{equation}
where 
\[ m^{(1)}(x) = F(x, m(x)),\] 
\[m^{(2)}(x) = \frac{\partial}{\partial x} F(x, m(x))
+ \left(\frac{\partial}{\partial m} F(x, m(x))\right)F(x, m(x)) \]
and so on.   Plugging these derivative expressions into 
(\ref{eqn:kernelReg}) yields a one-parameter nonlinear least-squares problem. 
That is, we minimize (\ref{eqn:kernelReg}) with respect to $m(x)$. 

Our goal, in this section, is to apply this technique to the estimation 
of $g(t)$ and $G(t)$.  


\subsection{Linear Scale Estimation}

For a given evaluation point $t$, where $t \in (a,b)$, the $p$th-order DE-constrained estimator for $g(t)$ is obtained by minimizing the local least-squares objective function
\[ \sum_{i=1}^n \left(y_i-\sum_{j=0}^p g^{(j)}(t)(t_i-t)^j/j!\right)^2K_h(t_i-t),\]
where $t_i$ are design points, the kernel function $K_h(t)$ is a symmetric probability density function scaled by the bandwidth $h$. The bandwidth $h$ satisfies the following condition:  $h\rightarrow 0$  and  $nh \rightarrow \infty$ as $n\rightarrow \infty.$

Using the differential equation $g'(t)=\lambda g^{\alpha}(t)$ in model (\ref{eqn:quasi1}), we obtain the $j^{th}$ derivative of $g(t)$:
\[g^{(j)}(t)=\lambda^p \left(\prod_{l=1}^{j}\{(l-1)\alpha-(l-2)\}\right) g^{j\alpha-p+1}(t).\]

We denote $\prod_{l=1}^{p}\{(l-1)\alpha-(l-2)\}$ as $\pi_{\alpha,p}$ in the sequal. With this notation, we write $g^{(p)}(t)=\lambda^p \pi_{\alpha,p} g^{p\alpha-p+1}(t)$.

\begin{align}
    &\doteq \sum_{i=1}^n \left\{y_i- g(t) - \sum_{p=1}^k \frac{1}{p!}(t_i-t)^p g^{(p)}(t)\right\}^2 K_h(t_i-t) \nonumber \\
    & = \sum_{i=1}^n \left\{y_i- g(t) - \sum_{p=1}^k \frac{1}{p!}(t_i-t)^p \lambda^p \pi_{\alpha,p} g^{p\alpha-p+1}(t)\right\}^2 K_h(t_i-t)
    \label{eqn:quasi2}
\end{align}

The $k^{th}$ degree DE-constrained estimator at $t$ is obtained by minimizing the weighted sum (\ref{eqn:quasi2}) with respect to the single parameter $g(t)$. The minimization is obtained by solving a weighted nonlinear least-squares problem, which is easily solved iteratively using the Gauss-Newton algorithm, given an appropriate initial guess. For example, the $2^{nd}$ degree DE-constrained estimator can be obtained using the local constant regression estimate as the starting value for the iteration. The local constant estimator for $g(t)$ handles sparse designs numerically better than higher-order local polynomial regression. Because it converges asymptotically to the true value at rate $O_p(n^{-2/5})$ under fairly general conditions, it can provide a good starting value for this iteration.

\subsection{Logarithmic Scale Estimation}

For estimation of $G(t)$, we use the $p^{th}$ derivative function of $G(t)$:
\[ G^{(p)}(t)=(p-1)!\lambda^p(\alpha-1)^{p-1}e^{p(\alpha-1)G(t)}\]
and apply the $k^{th}$ degree Taylor expansion for $G(t_i)$ in a sufficiently small neighborhood of $t$, we obtain:
\begin{align}
    &\quad \sum_{i=1}^n (\log(y_i) - G(t_i))^2K_h(t_i - t) \nonumber \\
    &\doteq \sum_{i=1}^n \left\{\log(y_i)- G(t) - \sum_{p=1}^k \frac{1}{p!}(t_i-t)^p G^{(p)}(t)\right\}^2 K_h(t_i-t) \nonumber \\
    & = \sum_{i=1}^n \left\{\log(y_i)- G(t) - \sum_{p=1}^k \frac{1}{p}(t_i-t)^p \lambda^p(\alpha-1)^{p-1}e^{p(\alpha-1)G(t)}\right\}^2 K_h(t_i-t)
    \label{eqn:quasilog2}
\end{align} 
By minimizing the weighted sum (\ref{eqn:quasilog2}) with respect to $G(t)$, we obtain the $k^{th}$ degree DE-constrained estimator for $G(t)$. Again, this minimization can be carried out using the Gauss-Newton algorithm. 

\section{Asymptotic Properties of the DE-Constrained Estimators}

In this section, we discuss the conditional asymptotic analysis of the $k^{th}$ degree DE-constrained estimator $\hat{g}_k(t)$. The following assumptions are made for the model (\ref{eqn:quasi1}). $g(t)$, the mean function, has a bounded and continuous $(k+1)^{th}$ derivative in a neighborhood of $x$. The design density, $f(t)$, is twice continuously differentiable and positive. And $K(\cdot)$, the kernel function, is a nonnegative, symmetric, and bounded PDF with compact support on the interval $[a,b]$. The kernel function satisfies $\int_{-\infty}^{\infty}K(w)dw=1$, $R(K)=\int K^2(w)dw < \infty$, and has finite moments up to sixth order. The rescaled kernel function is defined as $K_h(\cdot)=h^{-1}K(\cdot/h)$, inheriting the properties of the original kernel.

\subsection{Linear Scale Estimation}

Under the above assumptions, we have the following theorems about the asymptotic conditional bias and variance of the estimator $\hat{g}_k(t)$ for model (\ref{eqn:quasi1}).

\textbf{Theorem 1 (Asymptotic Conditional Bias)} For model (\ref{eqn:quasi1}), with $t \in (a+h, b-h)$, the $k^{th}$ degree DE-constrained estimator $\hat{g}_k (t)$ has asymptotic conditional bias
\begin{equation}
\mathrm{Bias}(\widehat{g}_k (t)|t_1,...,t_n) = \frac{1}{(k+1)!}g^{(k+1)}(t)h^{k+1}\mu_{k+1}+o_p(h^{k+1}),  \quad k \quad \text{odd},
\label{eqn:quasi5}
\end{equation}

where $\mu_{k+1}=\int w^{k+1}K(w)dw < \infty$, 

and when $k$ is even,
\begin{equation}
\resizebox{.9\hsize}{!}{
$\mathrm{Bias}(\widehat{g}_k (t)|t_1,...,t_n) =\frac{1}{(k+1)!}g^{(k+1)}(t)\left (\frac{\lambda [(k+1)\alpha -k]g^{\alpha-1}(t)}{k+2}+\frac{f'(t)}{f(t)}\right)h^{k+2}\mu_{k+2}+o_p(h^{k+2}),$}
\label{eqn:quasi6}
\end{equation}

where $\mu_{k+2}=\int w^{k+2}K(w)dw < \infty$.

In equation (\ref{eqn:quasi5}) and (\ref{eqn:quasi6}), $g^{(k+1)}(t)$ is the $(k+1)^{th}$ derivative of $g(t)$,
\[g^{(k+1)}(t)= \lambda^{k+1}\pi_{\alpha,k+1} g^{(k+1)\alpha-k}(t).\]

\textbf{Theorem 2 (Asymptotic Conditional Variance)} For model (\ref{eqn:quasi1}), with $t \in (a+h, b-h)$, the $k^{th}$ degree DE-constrained estimator $\hat{g}_k (t)$ has asymptotic conditional variance
\begin{equation}
\mathrm{Var}(\widehat{g}_k (t)|t_1,...,t_n) = \frac{\sigma^2R(K)}{nhf(t)}+o_p\left(\frac{1}{nh}\right).
\label{eqn:quasi7}
\end{equation}

The above theorems provide a tool to select the asymptotically optimal bandwidth for $\hat{g}_k (t)$ by minimizing the asymptotic mean squared error (AMSE).    

\textbf{Theorem 3 (Asymptotically Optimal Bandwidth)} Under the assumption of model (\ref{eqn:quasi1}), with $t \in (a+h, b-h)$, the asymptotically optimal bandwidths for the $k^{th}$ degree estimator $\hat{g}_k (t)$ are given by:
\begin{equation}
h_{o,k}^{2k+3}= \frac{\sigma^2R(K)((k+1)!)^2}{nf(t)\lambda^{2k+2}\pi_{\alpha,k+1}^2 g^{2(k+1)\alpha-2k}(t)(2k+2)\mu_{k+1}^2},
\label{eqn:oddh1}
\end{equation}
when $k$ is odd, and
\begin{equation}
\resizebox{.9\hsize}{!}{ $
h_{o,k}^{2k+5}= \frac{\sigma^2R(K)((k+1)!)^2}{nf(t)\lambda^{2k+2}\pi_{\alpha,k+1}^2 g^{2(k+1)\alpha-2k}(t)\left (\frac{\lambda [(k+1)\alpha -k]g^{\alpha-1}(t)}{k+2}+\frac{f'(t)}{f(t)}\right)^2(2k+4)\mu_{k+2}^2,} $}
\label{eqn:evenh1}
\end{equation}
when $k$ is even, where 
\begin{equation}    
g(t)=\{(1-\alpha)\lambda (t-t_0) +g(t_0)\}^{1/(1-\alpha)}, \quad  t_0 \quad \text{is the initial value of } t.
\end{equation}
which is the explicit solution to the differential equation in the model (\ref{eqn:quasi1}).

\textbf{Remark}: Theorem 3 provides the selection method for the asymptotically optimal bandwidths. The formulas (\ref{eqn:oddh1}) and (\ref{eqn:evenh1}) indicate the evaluation of the bandwidths for $k^{th}$ degree DE-constrained regression. In practice, the following formulas (\ref{eqn:oddh2}) and (\ref{eqn:evenh2}) are more useful if we obtain the start bandwidths when $k=0$ or $k=1$:
\begin{equation}
h_{o,k+2} =\Big(\frac{(k+3)(k+1)}{\lambda^4[(k+2)\alpha-(k+1)]^2[(k+1)\alpha-k]^2g^{4\alpha-4}(t)}h_{o,k}^{2k+3}\Big)^{1/(2k+7)}, 
\label{eqn:oddh2}
\end{equation}
when $k$ is odd, and 
\begin{equation}
\resizebox{.9\hsize}{!}{ $
h_{o,k+2} =\Big(\frac{(k+2)^3}{(k+4)\lambda^4[(k+1)\alpha -k]^2[(k+2)\alpha -(k+1)]^2g^{4\alpha -4}(t)}\frac{\left(\frac{\lambda [(k+1)\alpha -k]g^{\alpha-1}(t)}{k+2}+\frac{f'(t)}{f(t)}\right)^2}{\left(\frac{\lambda [(k+3)\alpha -(k+2)]g^{\alpha-1}(t)}{k+4}+\frac{f'(t)}{f(t)}\right)^2}h_{o,k}^{2k+5}\Big)^{1/(2k+9)}, $}
\label{eqn:evenh2}
\end{equation}
when $k$ is even.

For example, in a simulation study, we can use the formula (\ref{eqn:evenh1}) or the local constant regression to find $h_{o,0}$, that is, the asymptotically optimal bandwidth when $k=0$.  Then the asymptotically optimal bandwidth when $k=2$ can be obtained by the formula (\ref{eqn:evenh2}). Similarly, we use the formula (\ref{eqn:oddh1}) or the local linear regression to find $h_{o,1}$, that is, the asymptotically optimal bandwidth when $k=1$.  Then the asymptotically optimal bandwidth when $k=3$ can be obtained by the formula (\ref{eqn:oddh2}). Then, we can obtain the asymptotically optimal bandwidths for higher $k^{th}$ degree regressions step by step.

\subsection{Logarithmic Scale Estimation}

For the log scale model (\ref{eqn:quasilog1}), we can obtain analogous theorems on the asymptotic properties of the estimator $\hat{G}(t)$ by applying the above theorems and the $p^{th}$ derivative of $G(t)$,
\[G^{(p)}(t)=(p-1)!\lambda^p(\alpha-1)^{p-1}e^{p(\alpha-1)G(t)}.\]

Accordingly, we have a theorem about the asymptotic conditional bias of the estimator $\hat{G}_k(x)$ for model (\ref{eqn:quasilog1}) as follows.

\textbf{Theorem 4 (Asymptotic Conditional Bias)} For model (\ref{eqn:quasilog1}), with $t \in (a+h, b-h)$, the $k^{th}$ degree DE-constrained estimator $\hat{G}_k (t)$ has asymptotic conditional bias
\begin{equation}
\mathrm{Bias}(\widehat{G}_k (t)|t_1,...,t_n) = \frac{1}{(k+1)!}G^{(k+1)}(t)h^{k+1}\mu_{k+1}+o_p(h^{k+1}),  \quad k \quad \text{odd},
\label{eqn:quasi8}
\end{equation}

\noindent where $\mu_{k+1}=\int w^{k+1}K(w)dw < \infty$, and when $k$ is even,
\begin{equation}
\resizebox{.9\hsize}{!}{$
\mathrm{Bias}(\widehat{G}_k (t)|t_1,...,t_n) =\frac{1}{(k+1)!}G^{(k+1)}(t)\left (\frac{(k+1)\lambda(\alpha-1)e^{(\alpha-1)G(t)}}{k+2}+\frac{f'(t)}{f(t)}\right)h^{k+2}\mu_{k+2}+o_p(h^{k+2}),$}
\label{eqn:quasi9}
\end{equation}

\noindent where $\mu_{k+2}=\int w^{k+2}K(w)dw < \infty$.

In the equation (\ref{eqn:quasi8}) and (\ref{eqn:quasi9}), $G^{(k+1)}(t)$ is the $(k+1)^{th}$ derivative of $G(t)$,
\[G^{(k+1)}(t)= k!\lambda^{k+1}(\alpha-1)^{k} e^{(k+1)(\alpha-1)G(t)}.\]

The asymptotic conditional variance of the estimator $\hat{G}_k(t)$ has the same form as that given in Theorem 2. 

\textbf{Theorem 5 (Asymptotic Conditional Variance)} For model (\ref{eqn:quasilog1}), with $t \in (a+h, b-h)$, the $k^{th}$ degree DE-constrained estimator $\hat{G}_k (t)$  has asymptotic conditional variance
\begin{equation}
\mathrm{Var}(\widehat{G}_k (t)|t_1,...,t_n) = \frac{\sigma^2R(K)}{nhf(t)}+o_p\left(\frac{1}{nh}\right).
\label{eqn:quasi10}
\end{equation}

The asymptotically optimal bandwidth for $\hat{G}_k(t)$ can be derived using a similar approach to that described in Theorem 3. 

\textbf{Theorem 6 (Asymptotically Optimal Bandwidth)} For model (\ref{eqn:quasilog1}), with $t \in (a+h, b-h)$, the asymptotically optimal bandwidths for the $k^{th}$ degree estimator $\hat{G}_k (t)$ are given by:
\begin{equation}
h_{o,k}^{2k+3}= \frac{\sigma^2R(K)(k+1)!}{2nf(t)\lambda^{2k+2}(\alpha-1)^{2k} e^{2(k+1)(\alpha-1)G(t)}\mu_{k+1}^2},
\label{eqn:oddh3}
\end{equation}
when $k$ is odd, and
\begin{equation}
\resizebox{.9\hsize}{!}{$
h_{o,k}^{2k+5}= \frac{\sigma^2R(K)(k+1)^2}{nf(t)\lambda^{2k+2}(\alpha-1)^{2k}e^{2(k+1)(\alpha-1)G(t)}\left(\frac{(k+1)\lambda(\alpha-1)e^{(\alpha-1)G(t)}}{k+2}+\frac{f'(t)}{f(t)}\right)^2(2k+4)\mu_{k+2}^2},$}
\label{eqn:evenh3}
\end{equation}
when $k$ is even.

\noindent In equations \ref{eqn:oddh3} and \ref{eqn:evenh3}, 
\begin{equation}    
G(t)=(1/(1-\alpha))\log\{(1-\alpha)\lambda (t-t_0) +e^{(1-\alpha)G(t_0)}\}, 
\end{equation}
which is the explicit solution to the differential equation in the model (\ref{eqn:quasi1}) with initial value $t_0$.

In the following two sections, we evaluate the DE-constrained approach and compare it with other methods through two simulation studies and two real data analyses.

\section{Simulation Study on Mouse Tumor Growth Data}

In this section, we will consider the following two examples for the log scale model (\ref{eqn:quasilog1}), which pertains to two sets of control data from a chemotherapy trial in an animal experiment. The mouse tumor data (\citet{plume1993relative})  were collected on tumor volumes over time in mice. Tumor volume measurements were taken from a single mouse in Example 1 and 9 mice in Example 2. Times were recorded in days, and volumes were in cubic centimeters.  

\vspace{3mm}
\noindent
\textbf{Example 1: Models for One Mouse with Sparse Tumor Growth Data}
\vspace{3mm}

The single-mouse tumor data are provided in Table \ref{table:fullmouse1}. Times are recorded in days, and volumes are in cubic centimeters. To illustrate the performance of various local polynomial models on sparse data, we artificially removed some data points to create challenging estimation scenarios.

 \begin{table}[ht]
\centering
\scalebox{1}{
\begin{tabular}{ccc}
  \hline
  & time & volume \\ 
  \hline
1  &  21  & 0.05 \\
2  &  25  & 0.09 \\
3  &  28  & 0.22 \\
4  &  31  & 0.32 \\
5  &  33  & 0.61 \\
6  &  35  & 0.70 \\
7  &  38  & 0.90 \\
8  &  40  & 1.29 \\
9  &  42  & 1.77 \\
10 &  45 &  3.32 \\
  \hline  
\end{tabular}
}
\caption{The full set of observations on single-mouse tumor volume.} 
\label{table:fullmouse1}
\end{table}

Using the first-order differential equation in model (\ref{eqn:quasilog1}), the first-degree Taylor expansion for $G(t)$ in a sufficiently small neighborhood of $t$ gives
\begin{equation} 
\log(y_i) \doteq G(t) + \lambda \mbox{e}^{(\alpha - 1)G(t)}(t_i -  t) + \varepsilon_i. 
\label{eqn:1st}
\end{equation}

Furthermore, the second-degree Taylor expansion gives
\begin{equation} 
\log(y_i) \doteq G(t) + \lambda \mbox{e}^{(\alpha - 1)G(t)}(t_i - t) +
\frac{1}{2}\lambda^2(\alpha - 1)\mbox{e}^{2(\alpha - 1) G(t)}(t_i - t)^2 + 
 \varepsilon_i
 \label{eqn:2nd}
 \end{equation}
When implemented in the DE-constrained regression methodology, we refer to the models that apply the expansions (\ref{eqn:1st}) and (\ref{eqn:2nd}) as the first-degree and second-degree local quasi-growth models, respectively.


Since we do not know the true model for this data set, we use the local linear estimate fitted to the full data set as the reference ``truth''. The bandwidth was selected using the \verb!dpill! function \citep{ruppert1995effective} from the \textit{KernSmooth} package \citep{wand1995kernsmooth} in R, yielding $h = 2.38$. The standard deviation of the residuals was estimated as 0.089. 

For each simulation iteration, new observations were generated at the original design points $x_i$ ($i = 1, 2, \ldots, 10$) according to a normal distribution with mean $\widehat{G}_{ll}(x_i)$ and standard deviation 0.089. We created two sparse design scenarios: (1) removing observations 5, 6, 7, and 8 (moderately sparse), and (2) removing observations 4, 5, 6, 7, and 8 (highly sparse). For each scenario, 1000 simulated datasets were generated.

The competing methods are: local constant (NW), local linear (LL), local quadratic (LQ), first-degree local quasi-growth (DE1), second-degree local quasi-growth (DE2), and nonlinear least-squares (NLS) applied to the explicit solution of model (\ref{eqn:quasilog1}).

The DE-constrained methods require estimates of $\alpha$ and $\lambda$. From the explicit solution (\ref{eqn:soln}) to the differential equation and the log-transformation, we have
\[ G(t) \doteq \frac{1}{1-\alpha} (\log(1-\alpha) + \log(\lambda) + \log(t)), \quad \alpha \neq 1, \]
since $g(0)$ is necessarily very small in this application. This implies that the slope from simple linear regression of $\log(y)$ on $\log(t)$ estimates $1/(1-\alpha)$, which we use to obtain $\widehat{\alpha}$. Given this estimate, we then estimate $\lambda$ by applying nonlinear least-squares to the model
\[ y = \{(1-\widehat{\alpha}) (\lambda t)\}^{1/(1-\widehat{\alpha})}. \]

Table \ref{tab:param_summary} presents the five-point summaries of the parameter estimates across 1000 simulations for both sparse designs. The estimates show good stability, with $\alpha$ consistently centered around 0.819 and $\lambda$ around 0.151 for both scenarios.

For each simulated dataset, we computed the squared differences between $\widehat{G}(t_i)$ and $G_{\ell \ell}(x_i)$ at the removed observation points and averaged them across all removed points. Table \ref{logmodel} presents the mean of these average squared errors across 1000 simulations, along with standard errors in parentheses. Errors were calculated on both the log scale and the original scale (after exponentiating the fitted values).

The results demonstrate several important findings. First, local quadratic regression (LQ) experiences numerical instability in the highly sparse setting, with only 852 successful iterations out of 1000 when five points are removed. Second, NLS exhibits substantially larger errors than the local methods in both scenarios, suggesting that the global parametric model may be too rigid for these data.

Most notably, the DE-constrained methods demonstrate competitive performance compared to conventional local polynomial methods, particularly in sparse regions. In the moderately sparse scenario (removing 4 points), LL achieves the smallest log-scale error (0.0173), but DE1 is highly competitive with an error of only 0.0178—a difference of less than 3\%. Moreover, DE1 outperforms LL on the original scale (0.0141 vs 0.0149) and achieves the best performance in the highly sparse scenario on the original scale (0.0154).

It is worth noting that there is a slight inherent bias favoring LL in this comparison, since the reference "truth" was generated using a local linear fit. Despite this advantage, the DE-constrained estimators remain competitive, and in the highly sparse setting (removing 5 points), DE2 achieves the smallest log-scale error (0.0183), outperforming all conventional local polynomial methods.

These results indicate that DE-constrained methods can effectively guide kernel regression to produce satisfactory estimates even when data are sparse, by leveraging structural information from the underlying differential equation.

\begin{table}[ht]
\centering
\begin{tabular}{llrrrrr}
\hline
Design & Parameter & Min & Q1 & Median & Q3 & Max \\
\hline
Removing 5, 6, 7, 8 & $\alpha$ & 0.804 & 0.816 & 0.819 & 0.822 & 0.833 \\
 & $\lambda$ & 0.139 & 0.148 & 0.151 & 0.154 & 0.165 \\
Removing 4, 5, 6, 7, 8 & $\alpha$ & 0.804 & 0.816 & 0.819 & 0.822 & 0.833 \\
 & $\lambda$ & 0.100 & 0.148 & 0.151 & 0.154 & 0.165 \\
\hline
\end{tabular}
\caption{Five-point summary of $\alpha$ and $\lambda$ estimates across 1000 simulations for two sparse designs}
\label{tab:param_summary}
\end{table}

\begin{table}[ht]
\centering
\setlength{\tabcolsep}{4pt}  
\begin{tabular}{lcccc}
\hline
Method & \multicolumn{2}{c}{Removing 5,6,7,8} & \multicolumn{2}{c}{Removing 4,5,6,7,8} \\
\cline{2-3} \cline{4-5}
       & Log scale & Original scale & Log scale & Original scale \\
\hline
NW     & 0.2686 (0.0118) & 0.3019 (0.0146) & 0.4415 (0.0182) & 0.3549 (0.0159) \\
LL     & \textbf{0.0173 (0.0024)} & 0.0149 (0.0021) & 0.0341 (0.0044) & 0.0223 (0.0029) \\
LQ     & 0.0257 (0.0034) & 0.0219 (0.0029) & 0.0916 (0.0109) & 0.4900 (0.0581) \\
DE1    & 0.0178 (0.0024) & \textbf{0.0141 (0.0020)} & 0.0255 (0.0033) & \textbf{0.0154 (0.0022)} \\
DE2    & 0.0196 (0.0025) & 0.0229 (0.0028) & \textbf{0.0183 (0.0024)} & 0.0194 (0.0024) \\
NLS    & 0.1915 (0.0092) & 0.0482 (0.0051) & 0.4627 (0.0189) & 0.0635 (0.0063) \\
\hline
\end{tabular}
\caption{Average squared errors (standard errors in parentheses) for two sparse designs. Bold values indicate the best performance in each column. Note that DE1 (0.0178) is highly competitive with LL (0.0173) in the log scale with removing 5,6,7,8.}
\label{logmodel}
\end{table}

\vspace{3mm}
\noindent
\textbf{Example 2: Models for Multiple Mice with Simulated Tumor Growth Data}
\vspace{3mm}

While Example 1 demonstrated DE-constrained performance 
on a single trajectory under controlled sparse designs, tumor growth exhibits 
substantial heterogeneity across individuals. Example 2 evaluates whether 
group-level parameter estimates can effectively guide local regression for 
individual mice whose growth dynamics may deviate from the population average.

In this simulation study, we applied DE-constrained regression to other tumor growth data set collected on tumor volumes over time in 9 mice with 3 treatments: control, gold, and iodine. The first three and last two rows of the data set are in Table \ref{table:9mice}. We estimated $\lambda$ and $\alpha$ for each treatment group and then use these global estimates to perform first- and second-degree DE-constrained regression (DE1 and DE2 in short) on the tumor data for each mouse. We set up a bandwidth function using cross-validation and the \verb!thumbw! function in the \textit{locpol} package (\citet{cabrera2018package}) in R, and obtained the bandwidths used in the DE-constrained regression models for each treatment group. The detailed process is as follows. 

 \begin{table}[ht]
\centering
\scalebox{1}{
\begin{tabular}{ccccc}
  \hline
  & time & volume & mouse & treatment\\ 
  \hline
1 & 11 & 0.07 & 8 & c \\
2 & 13 & 0.18 & 8 & c \\
3 & 15 & 0.27 & 8 & c \\
... & ... & ... & ... & ...\\
74 & 30 & 3.05 & 66 & i \\
75 & 32 & 5.2 & 66 & i \\
  \hline  
\end{tabular}
}
\caption{The first three and last two observations in 9-mice tumor volume data, where c denotes the treatment Control and i denotes Iodine.} 
\label{table:9mice}
\end{table}

The simulation procedure began by estimating the global parameters $\lambda$ and $\alpha$ for each treatment group and extracting the time–volume data for each mouse. Using the estimated parameters, the differential equation solution was evaluated at each mouse’s observed time points to obtain the corresponding true volume values. Local linear regression was then fitted to the original data to estimate residual variability, which determined the noise level for the simulation. A total of 1000 simulated datasets were generated by adding random noise to the true values. For each dataset, optimal bandwidths were selected by cross-validation for six nonparametric and DE-constrained methods (NW, LL, LQ, LC, DE1, DE2), along with nonlinear least squares (NLS), and performance was assessed using the median absolute deviation (MAD). That is,
\[
\text{MAD} = \text{median}(|y_i - \hat{y}_i|),
\]
where \( y_i \) is the true model value, and \( \hat{y}_i \) is the fitted value for the corresponding data point.

Finally, the median and mean MAD values were summarized and visualized with boxplots to compare the seven methods, with lower MAD indicating better performance.

\begin{figure}[ht]
    \centering
    \includegraphics[width=\textwidth]{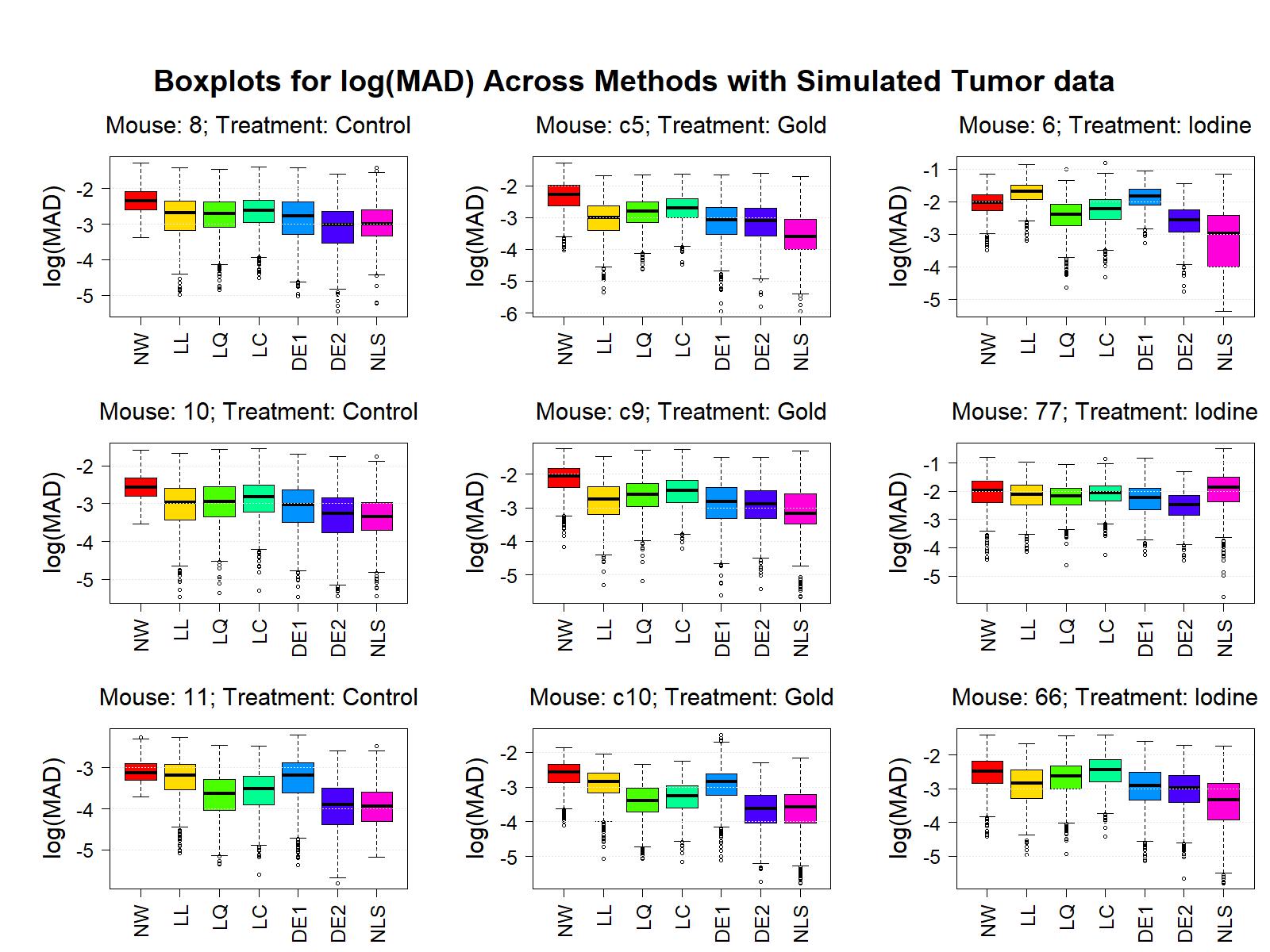}
    \caption{MAD (Median Absolute Deviation) distributions (on the log scale) for the estimation methods applied to simulated mouse tumor data based on 1000 simulation runs.}
    \label{9mice}
\end{figure}

With 1000 simulations, the MAD (Median Absolute Deviation) between $\hat{G(x_i)}$ and the true volume value $y_i$ is calculated. Figure \ref{9mice} shows the MAD distributions (on the log scale) for the following estimation methods:  NW (local constant), LL (local linear), LQ (local quadratic), DE1 (first-degree local quasi-growth), DE2 (second-degree local quasi-growth), and NLS (nonlinear least square). The overall boxplots for 9 mice show that DE2 approach achieves the best overall performance, and DE1 approach likewise demonstrates improved accuracy in estimating the mean function $G(x)$ relative to local polynomial regression methods. 

DE1 underperforms for some mice, specifically Control 11, Gold c10, and Iodine 6, when the local log-scale slope structure contradicts what the group-level $\hat\lambda$, $\hat\alpha$ impose. DE2 is more robust than DE1 not because it has more flexibility—both estimate only one parameter per grid point—but because its second-order Taylor expansion provides a more accurate local approximation of the true solution. The additional curvature constraint, while imposing more structure rather than less, stabilizes the local fit by capturing the natural concave-down shape of power-law growth that DE1 must approximate inefficiently through slope alone. This reduces residual variance and makes parameter estimation less sensitive to noise. Even when the imposed $\hat\lambda$ and $\hat\alpha$are slightly off from an individual mouse's true values, the coordinated bias in both the slope and curvature constraints produces a self-consistent local approximation that outperforms DE1's first-order approach. After DE2, NLS is the overall second-best approach except that it is very inaccurate for mouse Iodine 77, when this mouse data has a high $\sigma$ from residuals (0.298) and a clear non-monotone dip from time 18 to 22 with volume falls from 0.27 to 0.18. Local methods are robust because each estimate only depends on nearby points—a bad local fit at one region doesn't contaminate the others. NLS has no such protection.

Tables \ref{MADmean} and \ref{MADsd} provide numerical details of the MAD distribution on the original scale. 

\begin{table}[ht]
\centering
\scalebox{0.8}{
\begin{tabular}{rrrrrrrrrr}
  \hline
 & Control.8 & Control.10 & Control.11 & Gold.c5 & Gold.c9 & Gold.c10 & Iodine.6 & Iodine.77 & Iodine.66 \\ 
  \hline
NW & 102.29 & 81.78 & 47.20 & 105.79 & 128.20 & 76.81 & 138.36 & 147.19 & 87.81 \\ 
  LL & 72.41 & 56.20 & 42.78 & 54.71 & 69.75 & 59.05 & 186.38 & 129.62 & 65.21 \\ 
  LQ & 72.29 & 58.67 & 28.70 & 64.07 & 79.81 & 36.69 & 96.36 & 120.19 & 76.58 \\ 
  LC & 77.73 & 62.79 & 31.50 & 71.53 & 88.62 & 40.90 & 114.29 & 132.99 & 92.84 \\ 
  DE1 & 69.16 & 54.12 & 43.11 & 51.87 & 67.34 & 58.57 & 163.64 & 117.44 & 62.13 \\ 
  DE2 & 53.56 & 43.55 & 22.44 & 50.51 & 63.47 & 29.48 & 81.75 & 90.47 & 56.76 \\ 
  NLS & 59.08 & 41.35 & 21.80 & 36.89 & 57.92 & 30.90 & 60.68 & 169.20 & 41.23 \\ 
   \hline
\end{tabular}
}
\caption{Means of Median Absolute Deviation (MAD) multiplied by 1000 for the estimation methods applied to simulated mouse tumor data based on 1000 simulation runs.} 
\label{MADmean}
\end{table}

\begin{table}[ht]
\centering
\scalebox{0.8}{
\begin{tabular}{rrrrrrrrrr}
  \hline
 & Control.8 & Control.10 & Control.11 & Gold.c5 & Gold.c9 & Gold.c10 & Iodine.6 & Iodine.77 & Iodine.66 \\ 
  \hline
NW & 1.15 & 0.89 & 0.44 & 1.43 & 1.57 & 0.87 & 1.57 & 2.33 & 1.22 \\ 
  LL & 1.24 & 0.96 & 0.60 & 0.93 & 1.19 & 0.73 & 1.96 & 1.88 & 1.10 \\ 
  LQ & 1.17 & 0.96 & 0.46 & 0.95 & 1.20 & 0.54 & 1.37 & 1.63 & 1.14 \\ 
  LC & 1.16 & 0.96 & 0.45 & 0.96 & 1.26 & 0.56 & 1.54 & 1.72 & 1.27 \\ 
  DE1 & 1.27 & 0.98 & 0.66 & 0.95 & 1.24 & 0.84 & 1.75 & 1.95 & 1.11 \\ 
  DE2 & 0.99 & 0.83 & 0.41 & 0.92 & 1.15 & 0.49 & 1.18 & 1.41 & 1.01 \\ 
  NLS & 1.10 & 0.76 & 0.36 & 0.89 & 1.29 & 0.56 & 1.51 & 3.25 & 0.90 \\  
   \hline
\end{tabular}
}
\caption{Standard errors of mean for Median Absolute Deviation (MAD) multiplied by 1000 for the estimation methods applied to simulated mouse tumor data based on 1000 simulation runs.}
\label{MADsd}
\end{table}

\section{Data Analysis}

In this section, we examine the application of the DE-constrained method to actual mouse tumor growth data through two examples. The first example analyzes the data from a single mouse (Table \ref{table:fullmouse1}), while the second example considers data from nine mice (Table \ref{table:9mice}).

\vspace{3mm}
\noindent
\textbf{Example 3: Regression Fittings on One-Mouse Tumor Growth Data}
\vspace{3mm}

This example extends the study on sparse data from Example 1. To further examine the ability of the DE-constrained approach to handle severe data sparsity, we performed regression fittings on tumor growth data from a single mouse after removing 40\% of observations. The original dataset contained 10 measurements 
(Table \ref{table:fullmouse1}); we removed time points 5--8 (corresponding to days 33, 35, 38, 40), yielding a sparse dataset with only n = 6 observations spanning 24 days from day 21 to day 45. This reduction simulates realistic scenarios where data collection constraints limit temporal sampling, forcing 
estimation methods to infer growth patterns from minimal observations.

Parameter estimation via nonlinear least squares on the sparse data produced $\widehat{\alpha} =0.8731$ and $\widehat{\lambda} = 0.2035$, both indicating subexponential growth ($\widehat{\alpha} < 1$) consistent with the full dataset. The stability of parameter estimates despite the 40\% data loss suggests that 
the subexponential growth structure is robust and identifiable even with sparse sampling.

\begin{figure}[ht]
    \centering
    \includegraphics[width=\textwidth]{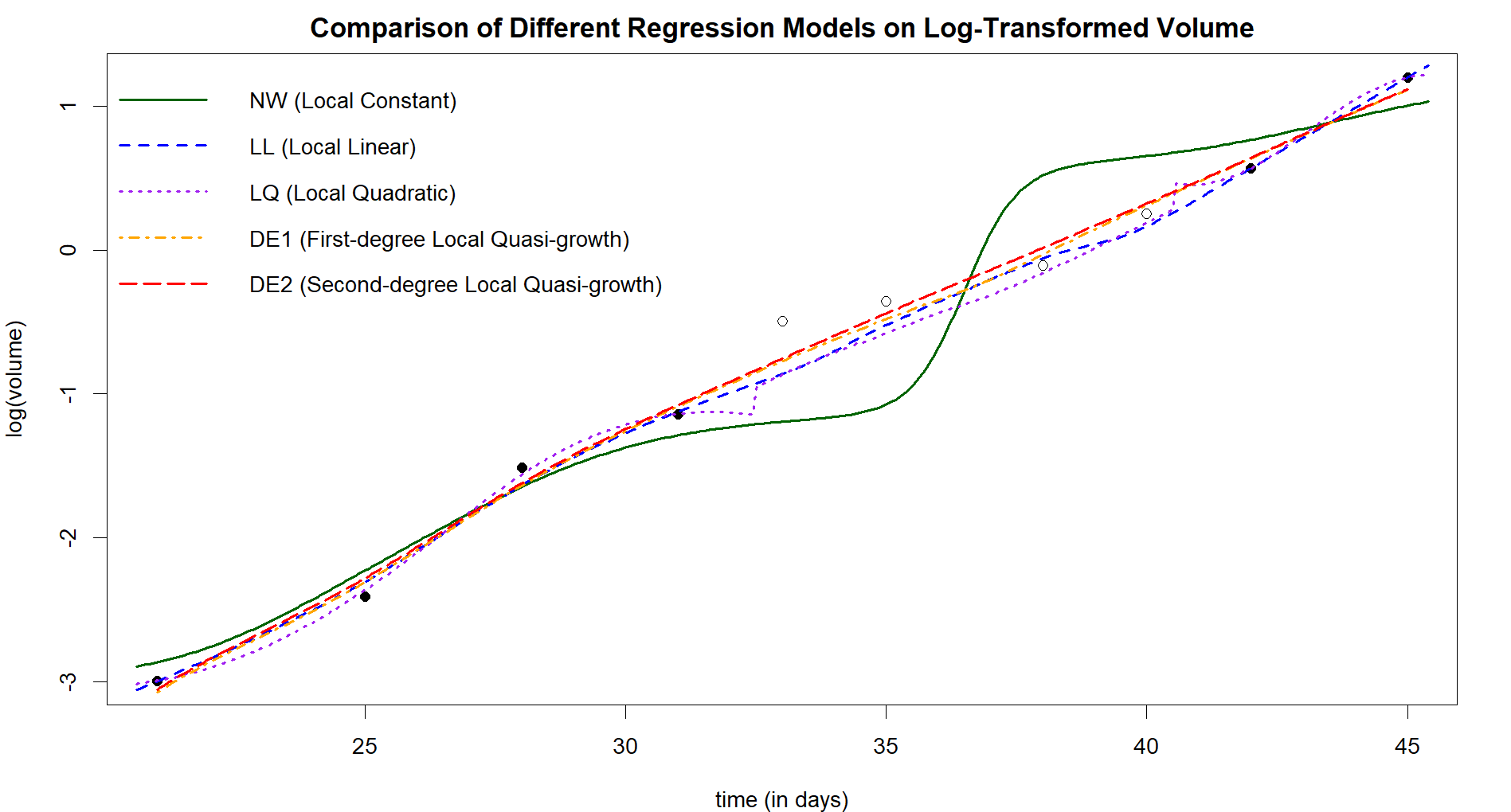}
    \caption{Fitted curves for different regression models on sparse tumor data removing points 5, 6, 7, and 8.}
    \label{fit}
\end{figure}

Figure \ref{fit} compares the fitted curves from various regression models on the sparse data. The comparison reveals a fundamental difference between conventional and 
DE-constrained approaches. Local polynomial regression methods—local constant (LC), local linear (LL), and local quadratic (LQ)—produce fits that exhibit substantial local oscillations between data points. These oscillations reflect the instability 
inherent in nonparametric methods when sample size is limited: with only 6 points distributed unevenly across the domain, local polynomial estimators lack sufficient data in some regions to stabilize their estimates.

In sharp contrast, the first- and second-degree DE-constrained local quasi-exponential growth models (DE1 and DE2) produce notably smoother, more stable fits. The smoothness 
arises from the incorporation of differential equation structure: rather than fitting arbitrary local polynomials, DE-constrained methods leverage the assumption that the true growth function satisfies the exponential growth differential equation (
 \ref{eqn:quasilog1}). This structural constraint effectively regularizes the estimation problem, reducing the degrees of freedom and providing stability when data are scarce.

The visual difference in Figure \ref{fit} illustrates a key principle: with sparse data, prior knowledge about the underlying mechanism (here, subexponential growth) provides crucial information that purely data-driven methods cannot access. The DE-constrained approaches successfully balance fidelity to the observed data points with adherence to the theoretical growth model, achieving both smoothness and interpretability in a 
way that conventional methods cannot match with limited observations.

This example demonstrates that DE-constrained methods are particularly valuable when data are sparse. The method's ability to produce stable, interpretable fits despite 40\% data removal highlights its potential utility in biomedical studies 
where practical constraints often limit the number of measurements per subject.

\vspace{3mm}
\noindent
\textbf{Example 4: DE-constrained Model for 9-Mouse Tumor Growth Data}
\vspace{3mm}

Example 4 presents a comprehensive application of the DE2 method (second-degree DE-constrained local quasi-growth model) to a multi-subject, multi-treatment dataset. The study included nine mice distributed across three treatment groups: 
Control (n = 3 mice), Gold nanoparticle treatment (n = 3 mice), and Iodine treatment (n = 3 mice). Each mouse had between 6 and 11 observations, with total sample sizes per treatment group ranging from 18 to 31 observations. This data structure—multiple 
subjects with sparse individual-level measurements—is typical of biomedical studies where practical constraints limit the frequency of measurements despite using multiple subjects.

Parameter estimation was performed at the treatment group level using pooled data. For each treatment group, we fitted the differential equation model (\ref{eqn:soln}) via nonlinear least squares, obtaining pooled estimates of $\widehat{\alpha}$ and 
$\widehat{\lambda}$ that apply to all mice within that group. All three treatment groups exhibited $\widehat{\alpha} < 1$, confirming subexponential growth dynamics across all conditions. The estimated growth parameters varied between treatments 
(Table \ref{tab:example4_params}), with differences in $\widehat{\lambda}$ suggesting that the treatments affected the rate of growth while maintaining the same fundamental growth mechanism.

\begin{figure}[ht]
    \centering
    \includegraphics[width=\textwidth]{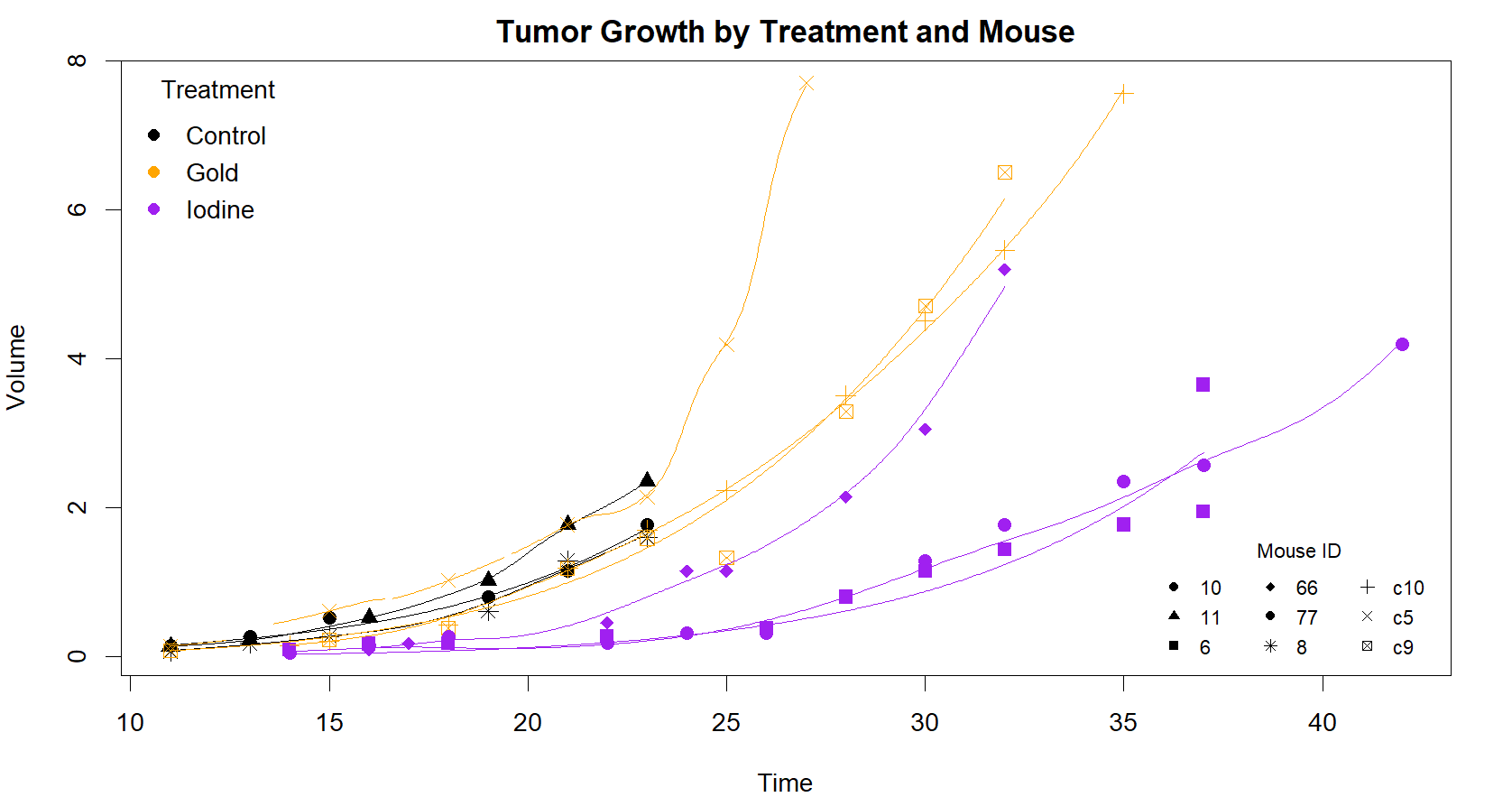}
    \caption{Second-degree DE-constrained local Quasi-exponential growth model on 9 mice tumor data for the 3 treatment groups: Control (black), Gold (Orange), and Iodine (Purple).}
    \label{DE2}
\end{figure}

Figure \ref{DE2} displays the DE2 fitting curves for all nine mice, organized by treatment group and colored by treatment. The DE2 fits are uniformly smooth across all mice and treatment groups, despite variation in individual sample sizes 
(6--11 observations per mouse) and different observed growth trajectories. The treatment effects are evident in the fitted curves, with Control mice showing moderate growth rates, Gold-treated mice exhibiting more rapid growth, and Iodine-treated mice displaying more restrained growth patterns.

The model fit quality, measured by RMSE and MAE (Table \ref{tab:example4_fit}), varies across treatment groups. The Control group shows superior fit quality 
(RMSE = 0.1958, MAE = 0.1443), while Gold and Iodine treatments exhibit larger residuals (Gold: RMSE = 0.9843, MAE = 0.5674; Iodine: RMSE = 0.7761, MAE = 0.4853), potentially reflecting greater within-treatment heterogeneity in growth dynamics. 
This pattern suggests that Control mice have more homogeneous growth behavior, while the treated groups show more individual variation in response to therapy.

The consistency of DE2 fits across subjects with different sample sizes validates the method's robustness to data sparsity in realistic multi-subject studies. Unlike conventional local polynomial methods that can become unstable with few observations, the DE-constrained approach leverages both the global treatment-level parameters (pooled $\widehat{\alpha}$ and $\widehat{\lambda}$) and local, subject-specific adaptation to produce reliable estimates. This balance between global structure and local flexibility is particularly valuable in applications where: (1) measurements are sparse within individual subjects, (2) multiple subjects are available for pooled estimation, and (3) both group-level inference and individual-level prediction 
are desired.

The nine-mouse example demonstrates the practical utility of DE-constrained methods for analyzing growth data in biomedical studies. The method successfully handles a realistic data structure with multiple subjects, multiple treatments, and sparse 
individual-level sampling. The smooth, interpretable fits across all mice establish DE-constrained estimation as a practical tool for tumor growth analysis and suggest applicability to other growth processes in biological and medical research.

\begin{table}[ht]
\centering
\begin{tabular}{lcc}
\hline
Treatment & $\widehat{\alpha}$ & $\widehat{\lambda}$ \\
\hline
Control   & 0.7440 & 0.2006 \\
Gold      & 0.6871 & 0.1748 \\
Iodine    & 0.6657 & 0.1158 \\
\hline
\end{tabular}
\caption{Pooled parameter estimates by treatment group for nine-mice tumor growth data. 
All three treatments exhibit $\widehat{\alpha} < 1$, confirming subexponential growth 
dynamics. The Control group shows the highest growth parameter ($\widehat{\lambda} = 0.2006$), 
while Iodine treatment exhibits the lowest ($\widehat{\lambda} = 0.1158$), suggesting 
differential treatment effects on growth rate.}
\label{tab:example4_params}
\end{table}

\begin{table}[ht]
\centering
\begin{tabular}{lccc}
\hline
Treatment & RMSE & MAE & $n$ obs \\
\hline
Control   & 0.1958 & 0.1443 & 18 \\
Gold      & 0.9843 & 0.5674 & 24 \\
Iodine    & 0.7761 & 0.4853 & 31 \\
\hline
\end{tabular}
\caption{Model fit quality (RMSE and MAE) for DE2 fits by treatment group. 
The Control group shows the best fit quality (lowest RMSE and MAE), while 
Gold and Iodine treatments show larger residuals, possibly reflecting greater 
heterogeneity in growth trajectories within these treatment groups.}
\label{tab:example4_fit}
\end{table}

\vspace{3mm}
\noindent
\textbf{Example 5: Fire Spread Dynamics}
\vspace{3mm}

While Examples 1--4 focused on tumour growth, the quasi-exponential model 
applies broadly to any subexponential growth process. We illustrate this 
with experimental fire spread data from \citet{ascmo-5-57-2019}, comprising 
6 replicated controlled burns where fire area was measured at half-day 
intervals. Table~\ref{tab:fire_data_params} summarises the data 
characteristics and parameter estimates for each replicate; unlike the 
sparse tumour growth data, these replicates are densely sampled with 
$n = 126$--$174$ observations each.

\begin{table}[ht]
\centering
\begin{tabular}{lccccc}
\hline
Replicate & $n$ & Time span (days) & Final area (m$^2$) & 
$\widehat{\alpha}$ & $\widehat{\lambda}$ \\
\hline
rep1 & 152 & [0, 75.5] & 3,877 & 0.4769 & 1.8320 \\
rep2 & 135 & [0, 67.0] & 4,447 & 0.3597 & 4.7659 \\
rep3 & 127 & [0, 63.0] & 3,363 & 0.4290 & 2.5648 \\
rep4 & 174 & [0, 86.5] & 3,264 & 0.5020 & 1.1728 \\
rep5 & 154 & [0, 76.5] & 6,359 & 0.3912 & 3.9777 \\
rep6 & 160 & [0, 79.5] & 5,708 & 0.3936 & 3.5628 \\
\hline
Mean & 150 & -- & 4,503 & 0.4254 & 2.9793 \\
SD   & --  & -- & --    & 0.0548 & 1.3624 \\
\hline
\end{tabular}
\caption{Summary of fire spread data and NLS parameter estimates for 6 
replicated controlled burns \citep{ascmo-5-57-2019}. Data are densely 
sampled at half-day intervals ($n = 126$--$174$), contrasting with the 
sparse tumour growth data in Examples 1--4. Parameters $\alpha$, $\lambda$ 
were estimated by fitting the log-scale explicit solution~\eqref{eqn:fire_nls} 
via nonlinear least squares. All replicates exhibit $\widehat{\alpha} < 1$, 
confirming subexponential growth dynamics.}
\label{tab:fire_data_params}
\end{table}

Specifically, parameters were estimated by fitting
\begin{equation}
\log(\text{area}) = \frac{1}{1-\alpha} \log\left\{(1-\alpha)\lambda t + 
e^{(1-\alpha)G_0}\right\}
\label{eqn:fire_nls}
\end{equation}
where $G_0 = \log(g(0))$ is the log-transformed initial area.

\begin{figure}[ht]
\centering
\includegraphics[width=\textwidth]{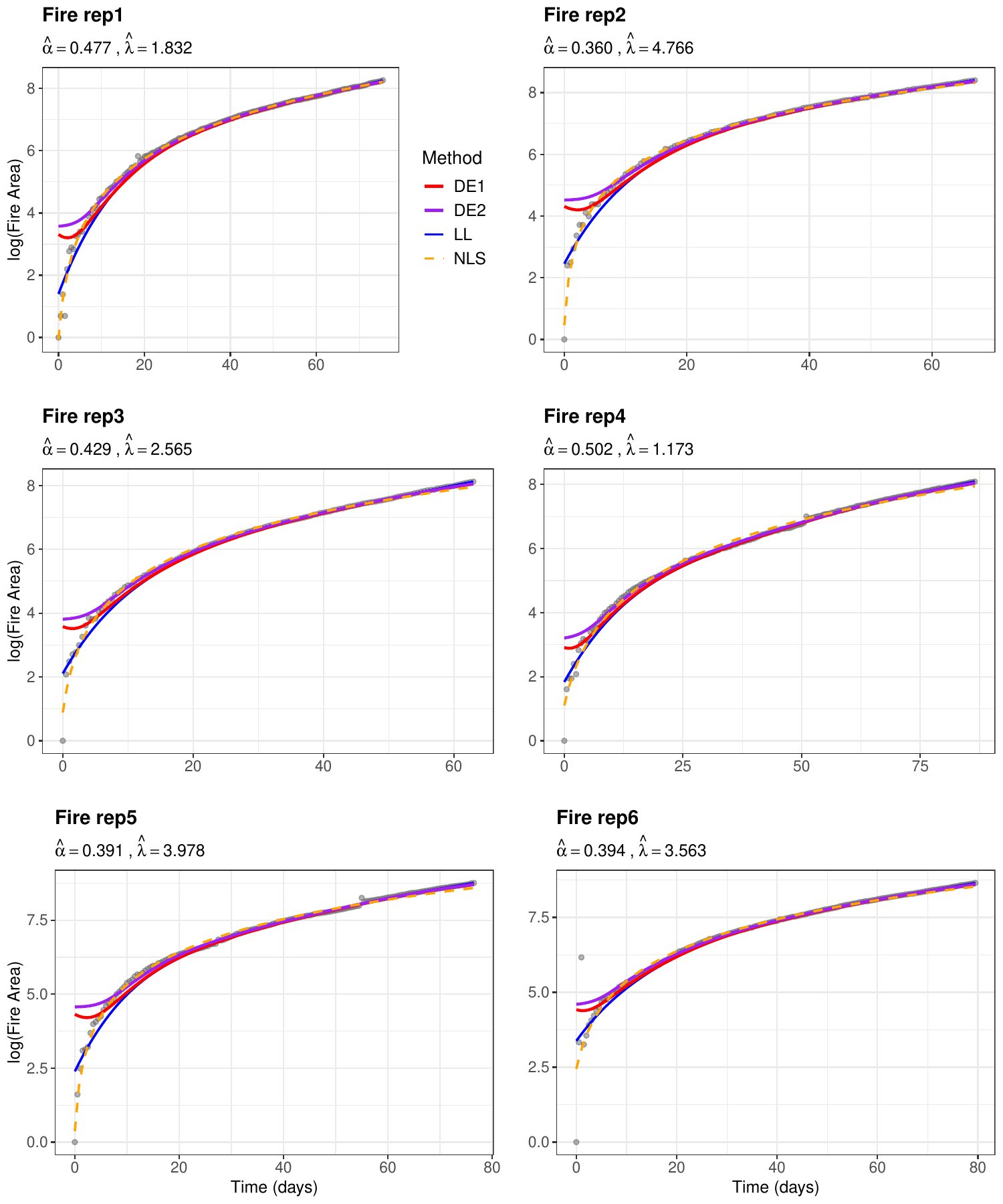}
\caption{Comparison of fitting methods across all six fire replicates on 
the log scale. Each panel shows observed data (gray points) and fitted 
curves from local linear (LL, blue), first-degree DE-constrained (DE1, 
red), second-degree DE-constrained (DE2, purple), and nonlinear least 
squares (NLS, orange dashed). Parameter estimates ($\widehat{\alpha}$, 
$\widehat{\lambda}$) are displayed for each replicate. Four methods shown 
for visual clarity; comprehensive comparison appears in 
Table~\ref{tab:fire_mse_asymptotic}.}
\label{fig:fire_all_replicates}
\end{figure}

We applied eight estimation methods to each replicate: NW, LL, LQ, 
DE1--DE4, and NLS. Figure~\ref{fig:fire_all_replicates} displays four 
representative methods (LL, DE1, DE2, NLS) for visual clarity; the 
DE-constrained methods produce smooth fits that respect the differential 
equation structure.

Table~\ref{tab:fire_mse_asymptotic} presents asymptotic MSE for seven 
methods, calculated as pointwise Bias$^2 +$ Var following Theorems~4 
and~5. NLS achieves the lowest mean MSE (0.0294), followed by DE3 (0.0469) 
and DE4 (0.0588). The large MSE values for LL reflect incompatibility with 
the DE-based asymptotic framework: bias here measures deviation from the 
theoretical DE solution, not from observed data. With dense data, NLS and 
higher-degree DE methods perform best; with sparse data (Examples 1--4), 
low-degree DE-constrained methods excel because structural constraints 
compensate for data scarcity.

\begin{table}[ht]
\centering
\small
\begin{tabular}{lccccccc}
\hline
Replicate & NW & LL & DE1 & DE2 & DE3 & DE4 & NLS \\
\hline
rep1 & 0.2000 & 2.7582 & 0.2399 & 0.2451 & 0.0639 & 0.0777 & \textbf{0.0106} \\
rep2 & 0.2249 & 4.2011 & 0.2081 & 0.2506 & 0.0562 & 0.0722 & \textbf{0.0118} \\
rep3 & 0.1248 & 1.2004 & 0.1661 & 0.1672 & 0.0451 & 0.0578 & \textbf{0.0182} \\
rep4 & 0.0628 & 0.3865 & 0.1214 & 0.0844 & 0.0310 & 0.0363 & \textbf{0.0197} \\
rep5 & 0.2088 & 4.2889 & 0.2188 & 0.2438 & 0.0565 & 0.0729 & \textbf{0.0131} \\
rep6 & 0.0683 & 0.4087 & 0.1013 & 0.0934 & \textbf{0.0287} & 0.0361 & 0.1030 \\
\hline
Mean & 0.1483 & 2.2073 & 0.1759 & 0.1807 & 0.0469 & 0.0588 & \textbf{0.0294} \\
SD   & 0.0727 & 1.7987 & 0.0559 & 0.0776 & 0.0145 & 0.0188 & 0.0362 \\
\hline
\end{tabular}
\caption{Asymptotic MSE (Bias$^2$ + Var) averaged across evaluation points 
for seven methods across six fire replicates, following Theorems~4 and~5. 
LQ is excluded as its MSE values (mean $= 16{,}865$) far exceed the range 
of the other methods, reflecting extreme sensitivity to the DE-based 
asymptotic criterion in the dense-data setting. NLS uses empirical MSE. 
Best method per replicate in bold.}
\label{tab:fire_mse_asymptotic}
\end{table}

We construct pointwise 95\% confidence bands using the asymptotic MSE 
decomposition from Theorems~4 and~5. The dense sampling ($n = 126$--$174$) 
supports reliable asymptotic inference here, unlike the sparse tumour growth 
settings of Examples 1--4. For each gridpoint $t$, the pointwise MSE is
\begin{equation}
\text{MSE}(t) = \text{Bias}^2(t) + \text{Var}(t),
\end{equation}
and the 95\% confidence bands are constructed as
\begin{equation}
\widehat{G}(t) \pm 1.96\sqrt{\text{MSE}(t)}.
\end{equation}
These are \emph{pointwise} bands: at each individual point $t$, the true 
function value lies within the interval with probability approximately 0.95; 
this is not a simultaneous statement over the entire curve.

\begin{figure}[ht]
\centering
\includegraphics[width=\textwidth]{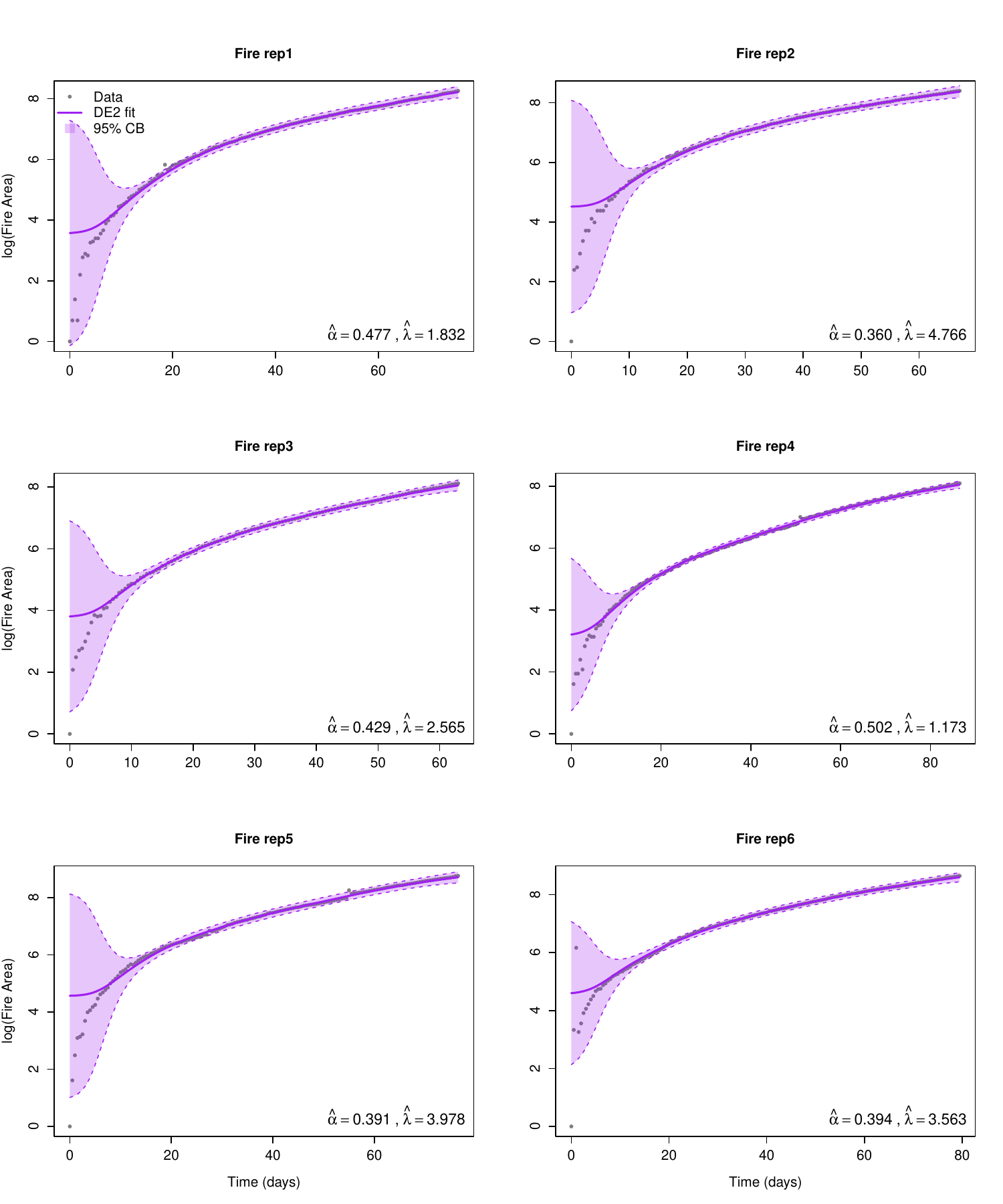}
\caption{95\% pointwise confidence bands for all six fire replicates using 
the DE2 estimator on the log scale. Each panel shows observed data (gray 
points), the DE2 fitted curve (purple line), and the shaded confidence 
region. Band widths vary across the domain, being narrower in high-density 
regions and wider near boundaries. These are pointwise bands: coverage 
probability applies at each individual point, not simultaneously for the 
entire curve.}
\label{fig:fire_all_reps_confidence_bands}
\end{figure}

Figure~\ref{fig:fire_all_reps_confidence_bands} displays the bands for all 
six replicates. Bands are narrower in high-density regions and wider near 
boundaries, reflecting the variance term $\sigma^2R(K)/(nhf(t))$. Empirical 
coverage ranges from 98.7\% to 99.4\% across replicates 
(Table~\ref{tab:confidence_bands_summary}), slightly exceeding the nominal 
95\% level, indicating the asymptotic approximation is moderately 
conservative. Table~\ref{tab:mse_decomposition_methods} shows that 
higher-degree DE-constrained estimators (DE3, DE4) achieve lower MSE 
through reduced bias at the cost of slightly increased variance, 
illustrating the classical bias-variance tradeoff.

\begin{table}[ht]
\centering
\begin{tabular}{lccc}
\hline
Replicate & $n$ & Coverage & Band Width \\
\hline
rep1 & 152 & 99.3\% & $\pm$0.461 \\
rep2 & 135 & 99.3\% & $\pm$0.480 \\
rep3 & 127 & 99.2\% & $\pm$0.391 \\
rep4 & 174 & 98.9\% & $\pm$0.265 \\
rep5 & 154 & 98.7\% & $\pm$0.467 \\
rep6 & 160 & 99.4\% & $\pm$0.303 \\
\hline
Mean & 150 & 99.1\% & $\pm$0.395 \\
\hline
\end{tabular}
\caption{Empirical coverage and average band width 
($1.96\sqrt{\text{MSE}}$) for 95\% pointwise confidence bands using the 
DE2 estimator. Mean coverage of 99.1\% slightly exceeds the nominal 95\% 
level, indicating the asymptotic approximation is moderately conservative. 
Wider bands for rep2 reflect smaller $n$ and larger estimated $\lambda$ 
(Table~\ref{tab:fire_data_params}).}
\label{tab:confidence_bands_summary}
\end{table}

\begin{table}[ht]
\centering
\begin{tabular}{lccccc}
\hline
Method & Avg $|\text{Bias}|$ & Avg Var & Avg MSE & Band Width & 
Bias$^2$/MSE \\
\hline
LL   & 0.6033 & 0.0014 & 4.2889 & $\pm$1.213 & 51.3\% \\
DE1  & 0.2172 & 0.0048 & 0.2188 & $\pm$0.497 & 39.4\% \\
DE2  & 0.1814 & 0.0058 & 0.2438 & $\pm$0.467 & 20.8\% \\
DE3  & 0.0932 & 0.0078 & 0.0565 & $\pm$0.316 & 19.4\% \\
DE4  & 0.1024 & 0.0089 & 0.0729 & $\pm$0.348 & 18.0\% \\
\hline
\end{tabular}
\caption{MSE decomposition for fire replicate 5 comparing estimation 
methods. Higher-degree DE methods reduce bias at the cost of increased 
variance, illustrating the classical bias-variance tradeoff.}
\label{tab:mse_decomposition_methods}
\end{table}

 \section{Discussion}
 
Information from differential equations can improve local polynomial regression estimates. The proposed methods are simple and have a low computational load without the requirement to solve the differential equation. We applied our proposed methods to mouse tumor growth data. The first-degree and second-degree DE-constrained regressions perform better than local constant and local quadratic regressions. The DE-constrained methods are also competitive with local linear regression when dealing with sparse regions.

In this paper, we focused on the first-order DE-constrained regression model, which involves a first-order differential equation. The model considered here is a first-order nonlinear differential equation with an explicit solution. The method employed in this paper does not require knowledge of this solution; it is generalizable to many other situations. In the future, we will 
explore higher-order models such as the second-order DE-constrained regression model.


\section*{Acknowledgements}

This research has been supported in part by a grant from the Natural Sciences and Engineering Research Council of Canada (NSERC).  

\bibliographystyle{apalike}
\bibliography{references}

\label{lastpage}

\end{document}